\documentclass[12pt]{iopart}

\usepackage{iopams}
\usepackage[pdftex]{graphicx}
\usepackage{color}
\usepackage{subfigure}
\usepackage{dcolumn}
\usepackage{bm}
\usepackage{epsf}
\usepackage{amssymb}
\usepackage{enumitem}
\usepackage{hyperref}
\usepackage[ansinew]{inputenc}
\usepackage{cite}
\usepackage[margin=20pt,font=footnotesize ,labelfont=bf]{caption}

\graphicspath{{./figures/}}

\newcommand{\be}{\begin{equation}}
\newcommand{\ee}{\end{equation}}
\newcommand{\bea}{\begin{eqnarray}}
\newcommand{\eea}{\end{eqnarray}}

\begin{document}

\setcounter{tocdepth}{2}

\title{Dynamics and control of fast ion crystal splitting in segmented Paul traps}

\author{H.~Kaufmann$^1$, T.~Ruster$^1$, C.~T.~Schmiegelow$^1$,  F.~Schmidt-Kaler$^1$, U.~G.~Poschinger$^1$}
\address{$^1$QUANTUM, Institut f\"ur Physik, Universit\"at Mainz,
  D-55128 Mainz, Germany}

\ead{h.kaufmann@uni-mainz.de}


\begin{abstract}
We theoretically investigate the process of splitting two-ion crystals in segmented Paul traps, i.e. the structural transition from two ions confined in a common well to ions confined in separate wells. The precise control of this process by application of suitable voltage ramps to the trap segments is non-trivial, as the harmonic confinement transiently vanishes during the process. This makes the ions strongly susceptible to background electric field noise, and to static offset fields in the direction of the trap axis. We analyze the reasons why large energy transfers can occur, which are impulsive acceleration, the presence of residual background fields and enhanced anomalous heating. For the impulsive acceleration, we identify the diabatic and adiabatic regimes, which are characterized by different scaling behavior of the energy transfer with respect to time. We propose a suitable control scheme  based on experimentally accessible parameters. Simulations are used to verify both the high sensitivity of the splitting result and the performance of our control scheme. Finally, we analyze the impact of trap geometry parameters on the crystal splitting process.
\end{abstract}

\tableofcontents

\section{Introduction}
Linear crystals of ions trapped in linear Paul traps have allowed for ground-breaking experiments in the fields of quantum computation, quantum simulation and precision measurements \cite{blatt2008entangled}. Segmented, micro-structured Paul trap arrays have been proposed as a future hardware platform for scalable quantum information experiments \cite{KIELPINSKI2002}. Small groups of ions are trapped separately from each other, such that precise manipulation of the qubits can be accomplished. Experimental  protocols then require ion shuttling operations, in addition to laser- or microwave-driven logic gates. Essential shuttling operations are splitting and merging of linear ion crystals.  It is important that they are fast on the typical timescale for quantum gates of 10-100$\mu$s, and in order to allow for gate operations or readout after the splitting, a low energy transfer is required.
Shuttling of trapped ions in segmented traps has been realized within a few oscillation cycles of the harmonic trap by time-dependent control of the trap voltages \cite{WALTHER2012,BOWLER2012}, at energy transfers below one motional quantum. Crystal splitting in a segmented trap was first demonstrated in Ref. \cite{ROWE2002}, at energy transfers of about 140 phonons within a splitting time of 10~ms. With optimizations, splitting has been included to the set of methods for quantum computing, e.g. for quantum teleportation \cite{BARRETT2004} and entanglement purification \cite{REICHLE2006}. Currently, the best reported result is a gain of about two vibrational quanta per ion at a time duration of 55~$\mu$s \cite{BOWLER2012}.
The experimental challenge for the control of this process is given by the fact that the harmonic part of the electrostatic trap potential has to change its sign during this process and therefore has to cross zero. This situation of weak confinement reduces the attainable speed and potentially increases the final motional excitation. In order to make the process more robust and faster, it is desirable to achieve a large quartic component of the axial trapping potential.\\
Trap geometries tailored to improve splitting performance were investigated in \cite{HOME2006}. Optimized geometry parameters for surface electrodes traps were derived in Ref. \cite{NIZAMANI2012}. In Ref. \cite{EBLE2010}, robust splitting operations on slow timescales were carried out by means of real-time observation of the ion positions and feedback on the segment voltages. \\
In this work, we analyze the splitting process with the aim of achieving low energy transfers in segmented miniaturized Paul traps. We reduce our analysis to the process of splitting ion crystals, as the process of merging ion crystals is merely the time reversed process. Furthermore, we restrict ourselves to the case of two ions. For splitting and merging processes with several ions, the general procedures and conclusions are still valid.\\
The manuscript is organized as follows:
In Sec. \ref{sec:Prereq}, we introduce the formalism for describing the electrostatic potentials during the splitting operations and the equilibrium positions of the ions, and we analyze the dependence of the equilibrium positions on the control parameters. In Sec. \ref{sec:Intricacies}, we give a detailed explanation of the possible reasons for high energy transfers. Based on these considerations, a procedure for the design of suitable voltage ramps is given in Sec. \ref{sec:voltageramps}. In Sec. \ref{sec:simu}, we analyze the performance of these ramps by numerical simulations. Finally, in Sec. \ref{sec:geom}, we compare typical examples for trap geometries and discuss the implication for ion splitting.

\section{Prerequisites}
\label{sec:Prereq}
\subsection{Electrostatic trap potentials}
\label{sec:potentials}
\begin{figure}[h!]
\centering
\includegraphics[width=\textwidth]{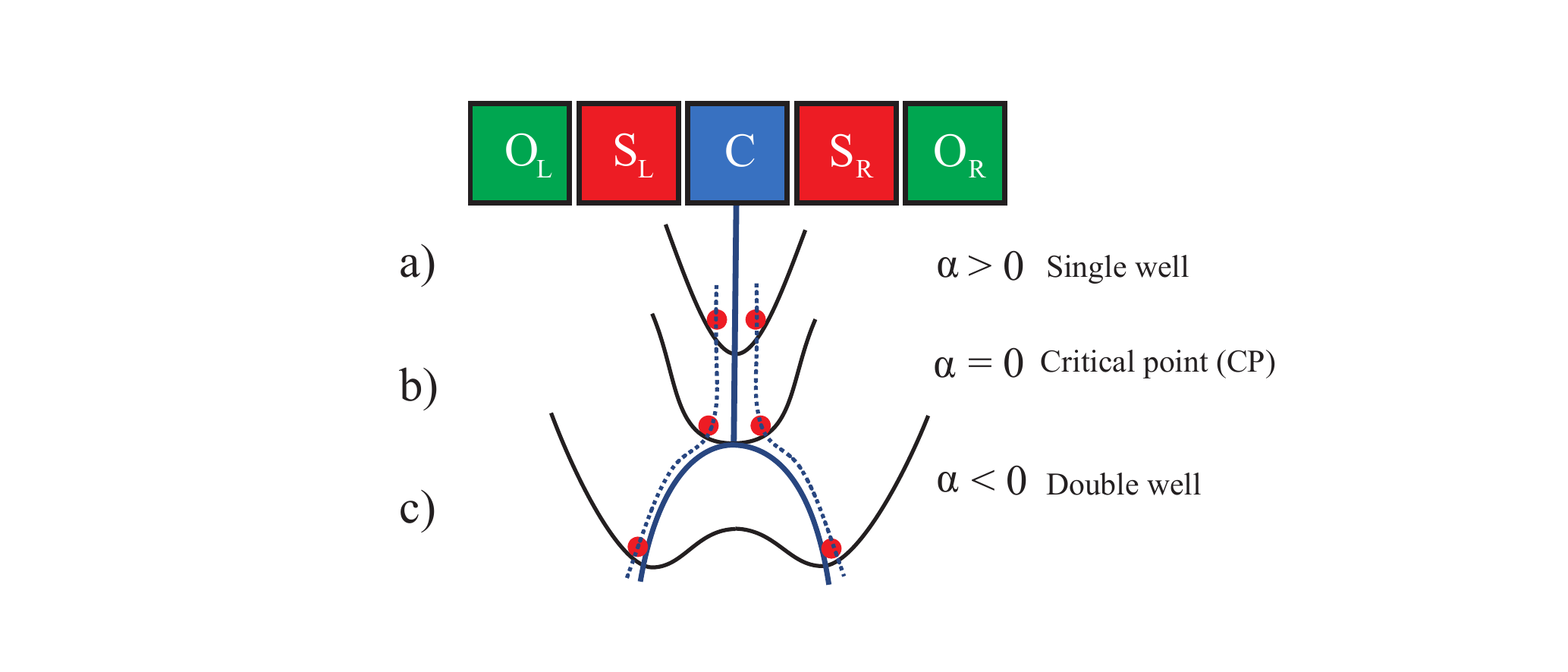}
\caption{The process of ion crystal splitting. It is shown schematically how two ions are moved from the initial center segment $C$ to different destination segments $S_{R,L}$ by changing a confining electrostatic potential from \textbf{a)} a strong harmonic confining potential ($\alpha>0$) via \textbf{b)} a predominantly quartic potential ($\alpha\approx 0$) to \textbf{c)} a double-well potential ($\alpha<0$). The external potential is determined by the voltages applied to the respective electrodes. The equilibrium positions are sketched as dashed lines. The outer electrodes $O$ facilitate the splitting process by increasing the transient quartic confinement and offer the possibility to cancel a possible axial background field by application of a differential voltage. The color coding of the segments and the corresponding voltages is used throughout the manuscript.}
\label{fig:intro}
\end{figure}

We desire to split a two-ion crystal residing at center segment $C$ along the trap axis $x$, to obtain two ions stored in separated potential wells at the position of the splitting segments $S$ neighboring $C$, see Fig. \ref{fig:intro}.
Note that we consider only the spatial dimension along the trap axis, as we assume that tight radial confinement persists throughout the process and the ions are always located on the rf node of the trap. Typical distances between segments range between 50 and 500~$\mu$m, while the initial ion distance is 2-4~$\mu$m.  The total external electrostatic potential along the trap axis can be written as
\be
\Phi(x)\approx\beta~x^4+\alpha~x^2+\gamma~x
\label{eq:taylorpotential}
\ee
where the coefficients $\alpha,\beta,\gamma$ are given by the the trap geometry and the voltages applied to the trap segments. This Taylor approximation is valid as long as the the ions are located sufficiently close to  $x=0$, which is the center of the $C$ segment.
Throughout the splitting process, the external potential is changing from a single well potential $\alpha_i>0$ to a double well potential $\alpha_f<0$, crossing the critical point (CP) at $\alpha=0$. Note that $\beta>0$ is required to guarantee confinement at $\alpha\leq 0$. The approximation of Eq. \ref{eq:taylorpotential} holds for $\alpha\geq 0$ and for $\alpha\lesssim0$ as long as the separation of the two potential wells is small compared to the width of segment $C$. When the distance of the ions from the center of the $C$ segment becomes comparable to the width of the segment, anharmonic terms of order $>4$ contribute significantly to the total potential. These are not taken into account here since the outcome of the splitting process is determined around the CP, as will be pointed out in the following sections. Furthermore, beyond the CP, the distance of the separated wells is still increasing monotonically for decreasing $\alpha$ as long as the variation $\beta$ is sufficiently small, and the corresponding trap frequencies in these wells are monotonically increasing. For studies which require precision beyond the CP, the higher order terms can be taken into account numerically.
A cubic term does not contribute to the potential if the trap is sufficiently symmetric along the trap axis.\\
Including Coulomb repulsion, the total electrostatic potential of a two-ion crystal at a center-of-mass position $x_0$ and distance $d$ is given by
\be
\Phi_{tot}(x_0,d)=\Phi(x_0+d/2)+\Phi(x_0-d/2)+\frac{\kappa}{d},
\label{eq:totalpotential}
\ee
with $\kappa=e/4\pi\epsilon_0$. At the CP, the harmonic confinement vanishes, and a weak residual confinement is maintained by the interplay between Coulomb repulsion and quartic part of the external potential. It is therefore desirable to maximize $\beta$ at the CP. For a given trap geometry, the attainable $\beta$ is limited by the voltage range which can be applied to the trap electrodes \footnote{The maximum voltage is ultimately limited by the electric breakdown threshold. In practice, as precisely controlled time-dependent voltage waveforms are to be applied to the trap segments, the voltage range will be determined by the electrical design, where one faces a trade-off between voltage range and output bandwidth \cite{BAIG2013,BOWLER2013}.}.
The coefficients of the potential Eq. \ref{eq:taylorpotential} are given by the segment bias voltages and the electrostatic properties of the trap:
\bea
\alpha&=&U_C~\alpha_C+U_S\alpha_S+U_O\alpha_O \label{eq:alphadef} \\
\beta&=&U_C~\beta_C+U_S\beta_S+U_O\beta_O \label{eq:betadef} \\
\gamma&=&\Delta U_S\gamma_S+\Delta U_O\gamma_O+\gamma'\label{eq:gammadef}
\eea
An offset parameter $\gamma'$ is introduced for taking trap non-idealities -- leading to a symmetry breaking force along the trap axis --  into account, see Sec. \ref{sec:uncomptilt}. In contrast to the symmetric quadratic and quartic contributions, the asymmetric tilt potential is controlled by the differential voltages $\Delta U_{S,O}$ between the corresponding left and right electrodes of the respective pair. The segment coefficients are given by Taylor expansions of the standard potentials $\phi_n(x)$, which are the dimensionless electrostatic potentials along the trap axis if a +1V bias is applied to segment $n$ and all other segments are grounded \cite{BLAKESTAD2011,SINGER2010}:
\be
\phi_{n,m}(x)=\phi_n\vert_{x_0^{(m)}}+\phi_n'\vert_{x_0^{(m)}}\delta x+\frac{1}{2}\phi_n''\vert_{x_0^{(m)}}\delta x^2+\frac{1}{24}\phi_n^{(4)}\vert_{x_0^{(m)}}\delta x^4+\mathcal{O}\left(\delta x^6\right).
\ee
with $\delta x=x-x_0^{(m)}$, i.e. the Taylor expansions are carried out at center of segment $m$, $x_0^{(m)}$. The coefficients for Eqs. \ref{eq:alphadef},\ref{eq:betadef},\ref{eq:gammadef} are obtained for $m=C,n=C,S,O$:
\be
\alpha_n=\frac{1}{2}f_n\phi_{n,C}''(0), \qquad \beta_n=\frac{1}{24}f_n\phi_{n,C}^{(4)}(0), \qquad \gamma_n=f_n\phi_{n,C}'(0),
\label{eq:alphabetafromPhi}
\ee
with $f_{C}=1$ and $f_{S,O}=2$ accounting for two $S,O$ segments acting symmetrically at $x=0$. Note that $\gamma_C=0$ by definition. \\
In the following, for numerical calculations, we use the specific geometry parameters of a three dimensional microstructured segmented ion trap A as detailed in Sec.\ref{sec:geom}.  There, other traps and their geometry parameters are listed and analyzed as well.

\subsection{Equilibrium positions}
\label{sec:eqpos}

\begin{figure}[h!]
\centering
\includegraphics[width=1.0\textwidth]{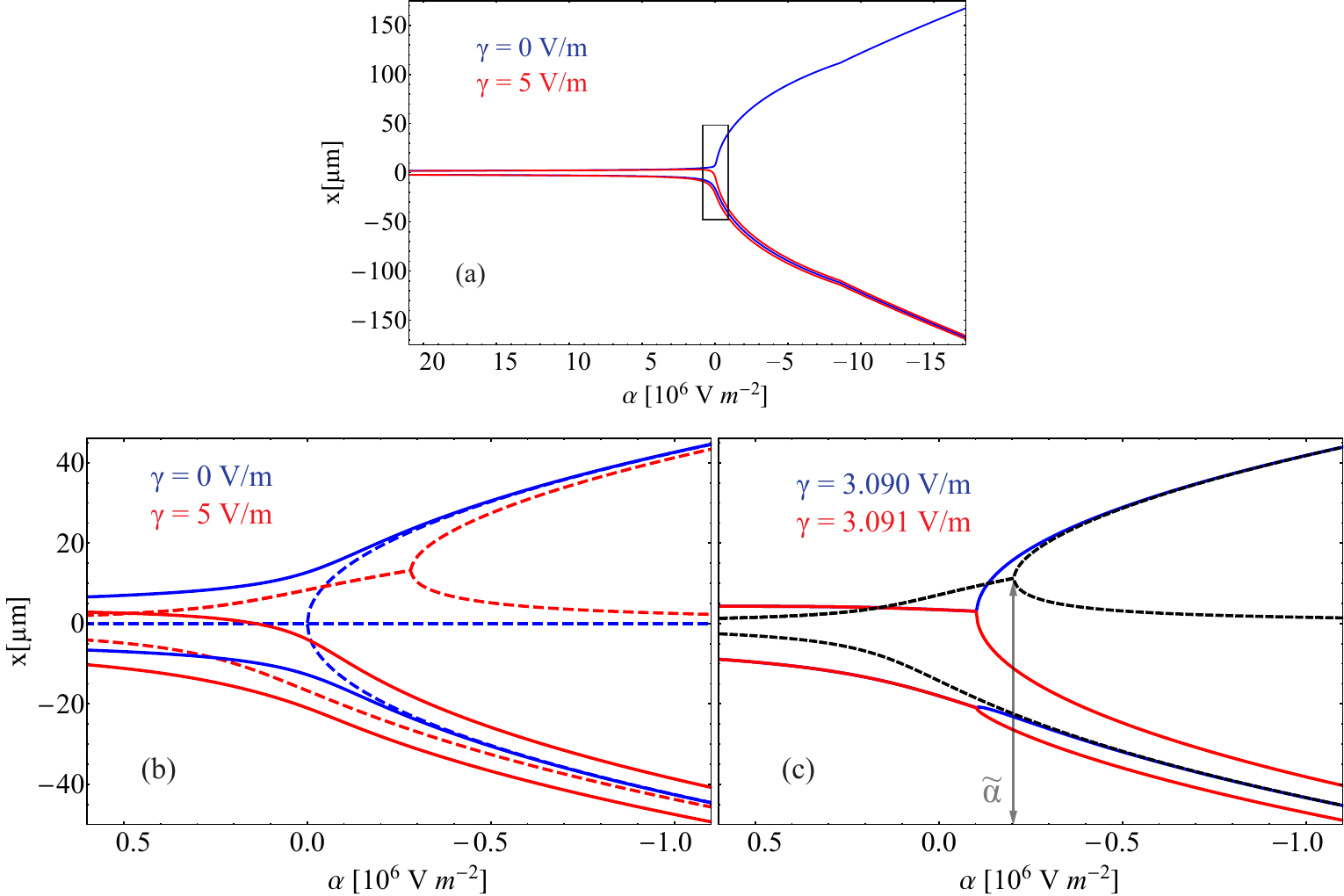}
\caption{Ion equilibrium positions near the critical point. \textbf{a)} shows the equilibrium positions  versus the harmonic parameter $\alpha$. In the case of a perfectly compensated tilt (blue), the ions separate symmetrically, in the case of a large tilt (red), both ions move towards one side. Panel \textbf{b)} shows a close-up around the critical point for the same tilt parameters. Additionally, the minima of the external potential are shown (dashed). In panel \textbf{c)}, we display equilibrium positions and potential minima for tilt parameters slightly below (blue) and above (red) the critical tilt parameter. In contrast to the corresponding curves in b), the equilibrium positions exhibit cusps which lead to strongly enhanced acceleration. }
\label{fig:eqnposnstogether}
\end{figure}

We consider two ions of mass $m$ and charge $e$, with their equilibrium positions given by the center-of-mass $x_0$ and the equilibrium distance $d$:
\be
x_{L,R}=x_0\pm d/2,
\ee
determined by minimizing of the total electrostatic potential Eq. \ref{eq:totalpotential}. The confinement is characterized by the local trap frequency, which is given by the curvature of of the external potential at the ion positions:
\be
\omega=\sqrt{\frac{e}{m}\Phi''(x_{L,R})}.
\ee
The extremal points of the external potential  Eq. \ref{eq:taylorpotential} are given by
\bea
x_0^{(0)}&=&\frac{\alpha}{3^{1/3}\zeta}-\frac{\zeta}{2\cdot3^{2/3}\beta}  \\
x_0^{(\pm)}&=&\frac{(i\sqrt{3}\pm 1)\alpha}{2\cdot3^{1/3}\zeta}+\frac{(1\mp i\sqrt{3})\zeta}{4\cdot3^{2/3}\beta} \\
\label{eq:potentialextrema}
\eea
where
\be
\zeta(\alpha,\beta,\gamma)=\left(9\beta^2\gamma+\sqrt{3}\sqrt{8\alpha^3\beta^3+27\beta^4\gamma^2}\right)^{1/3}.
\ee
Initially, at $\alpha=\alpha_i$, the confining harmonic part of the external potential and the Coulomb repulsion are dominant, thus we can neglect the quartic potential. The trap frequency is then given by $\omega^2=2\alpha e/m$ at an ion distance of $d=\left(\kappa/\alpha\right)^{1/3}$.  At the CP, $\alpha =0$, and without tilt, $\gamma=0$, the ion distance is determined by quartic confinement and Coulomb repulsion:
\be
d_{CP}=\left(2\kappa/\beta\right)^{1/5}.
\label{eq:dcp}
\ee
The Coulomb repulsion pushes the ions away from the trap center (where the curvature of the external potential vanishes), such that a residual harmonic confinement persists because of the quartic term. The minimum trap frequency during the splitting process is thus given by \cite{HOME2006}
\be
\omega_{CP}=\beta^{3/10}\left(3e/m\right)^{1/2}\left(2\kappa\right)^{1/5}.
\label{eq:omegamin}
\ee
Near the CP, the equilibrium distance can be computed from a perturbative expression up to second order:
\be
d(\alpha)\approx d_{CP}-\frac{1}{5}\left(\frac{16}{\beta^4\kappa}\right)^{1/5}\alpha+\frac{2}{25}\left(\frac{4}{\beta^7\kappa^3}\right)^{1/5}\alpha^2,
\label{eq:dpert}
\ee
for $\vert\alpha\vert\ll\beta d_{CP}^2$ and $\vert\alpha\vert\ll\kappa d_{CP}^{-3}$.\\
The center-of-mass position of the ion crystal near the critical point to first order in the tilt parameter $\gamma$ is:
\be
x_0(\alpha,\gamma)\approx \gamma \left(-\frac{1}{3\cdot 2^{2/5} \beta^{3/5} \kappa^{2/5}}-\frac{2^{1/5}}{45\cdot\beta^{6/5} \kappa^{4/5}}\alpha+\frac{26\cdot2^{4/5}}{675\beta^{9/5} \kappa^{6/5}}\alpha^2\right)
\label{eq:x0pert}
\ee
If the ions are sufficiently separated, $\alpha\ll 0$, the Coulomb repulsion can be neglected and the equilibrium positions approximately coincide with the extrema of the external potential:
\be
d_f=\sqrt{-2 \alpha_f/\beta}
\ee
and the final trap frequency is given by $\omega_f^2=-4\alpha_f e/m$.

\subsection{Critical tilt value}
A static background force along the trap axis can to keep the ions confined in one common potential well throughout the splitting process. We make use of the external potential minima Eqs. \ref{eq:potentialextrema} to obtain an estimate for the tilt parameter $\tilde{\gamma}$, beyond which the splitting ceases to work. In the following, we assume $\gamma>0$.

\begin{figure}[ht!]
\centering
\includegraphics[width=8.5cm]{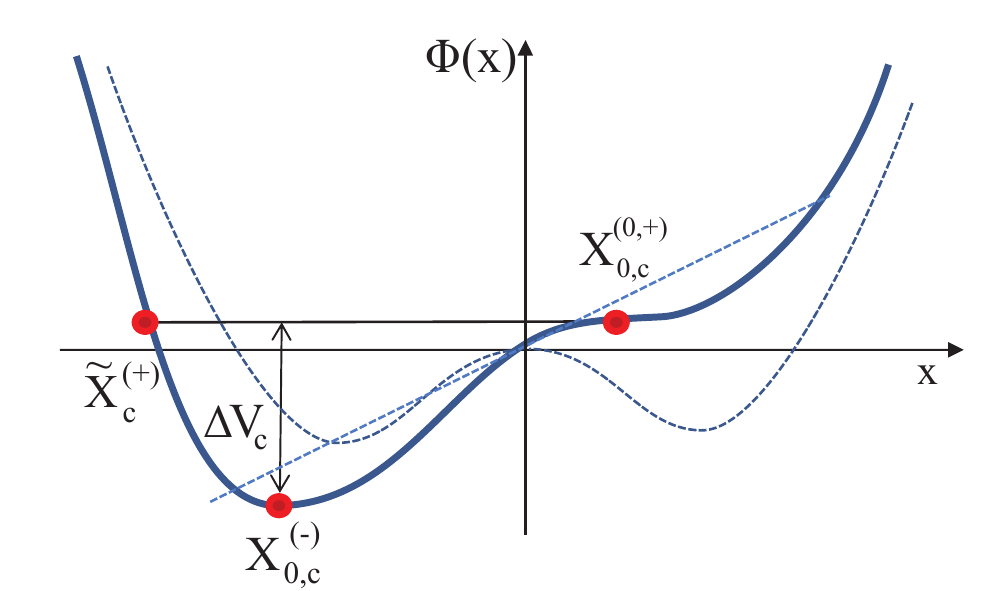}
\caption{Critically tilted potential, see text such that the Coulomb repulsion fails to push the right ion across the saddle point. }
\label{fig:gammacritpotential1}
\end{figure}
In the presence of a nonzero potential tilt, an imperfect bifurcation occurs, i.e. the second potential well opens up at $\tilde{\alpha}<0$, see Fig. \ref{fig:eqnposnstogether} c). We obtain a scaling law for $\tilde{\gamma}$ by calculating at which tilt parameter the original potential well is deep enough to keep both mutually repelling ions confined, see Fig. \ref{fig:gammacritpotential1}. The saddle point where the second potential well opens can be found by solving $x_{0,c}^{(0)}=x_{0,c}^{(+)}$ for $\tilde{\alpha}$, yielding
$\tilde{\alpha}=-{\textstyle\frac{3}{2}}\beta^{1/3}\vert\gamma\vert^{2/3}$. From this we obtain its position \footnote{For $\gamma\geq 0$, $x_0^{(0)}$ corresponds to the left potential minimum which always exists, and for $\alpha<\tilde{\alpha}<0$, $x_0^{(+)}$ corresponds to the right potential minimum and $x_0^{(-)}$ corresponds to the maximum of the separation barrier. By contrast, for $\gamma<0$, $x_0^{(0)}$ corresponds to the right potential minimum, and for $\alpha<0<\tilde{\alpha}$, $x_0^{(+)}$ corresponds to the left minimum.} to be $x_{0,c}^{(+,0)}={\textstyle\frac{1}{2}}\left(\gamma/\beta\right)^{1/3}$. At $\tilde{\alpha}$, the left potential minimum is located at twice the distance from the origin $x_{0,c}^{(-)}=-\left(\gamma/\beta\right)^{1/3}$. The potential attains the same value as on the saddle point $V(x_{0,c}^{(+,0)})$ at the position $\tilde{x}_{c}^{(+)}=-{\textstyle\frac{3}{2}}\left(\gamma/\beta\right)^{1/3}$. The depth of the potential well defined by the saddle point when the right well opens is therefore
\be
\Delta V_{c}=V(x_{0,c}^{(-)})-V(x_{0,c}^{(+,0)})=\frac{27}{16}\left(\frac{\gamma^4}{\beta}\right)^{1/3}.
\label{eq:critwelldepth}
\ee
We can now define a criterion which determines whether the ions are actually separated by comparing the Coulomb potential to the depth of the initial well at the CP, Eq. \ref{eq:critwelldepth}: If the Coulomb repulsion pushes the right ion beyond the saddle point $x_{0,c}^{(+,0)}$, it will end up in the right potential well, otherwise the two ions will stay in the left well. Thus, the Coulomb energy at an ion distance of $x_{0,c}^{(+,0)}-\tilde{x}_{c}^{(+)}$ has to be larger than the well depth $\Delta V_{c}$.
These considerations lead to a critical tilt value of
\be
\tilde{\gamma} < \pm ~C_{\gamma}\left(\kappa^3 {\beta^2}\right)^{1/5}.
\label{eq:scalingofgammac}
\ee
Despite the fact that the situation depicted Fig. \ref{fig:gammacritpotential1} does not actually occur, as the external force at the saddle point vanishes and therefore cannot balance the Coulomb force, the obtained scaling behavior is confirmed by numerical calculations, revealing a prefactor of $C_{\gamma}=$1.06.\\
The result Eq. \ref{eq:scalingofgammac} enables us to determine the required degree of precision by which the background axial field has to be corrected. For this calculation, only the geometry parameters $\beta_{C,S,O}$ are needed. Furthermore, the sensitivity decreases as $\beta^{2/5}$, which directly characterizes the gain in robustness when the accessible voltage range is enhanced. For trap A (Sec. \ref{sec:geom}), we derive a value of $\tilde{\gamma}\approx 3V/m$, corresponding to the requirement to set $\Delta U_O$ more accurately than about 9~mV. \\

\section{Intricacies of crystal splitting}
\label{sec:Intricacies}
\subsection{Impulsive acceleration at the critical point}
\label{sec:accatCP}

\begin{figure}[h!]
\centering
\includegraphics[width=1.0\textwidth]{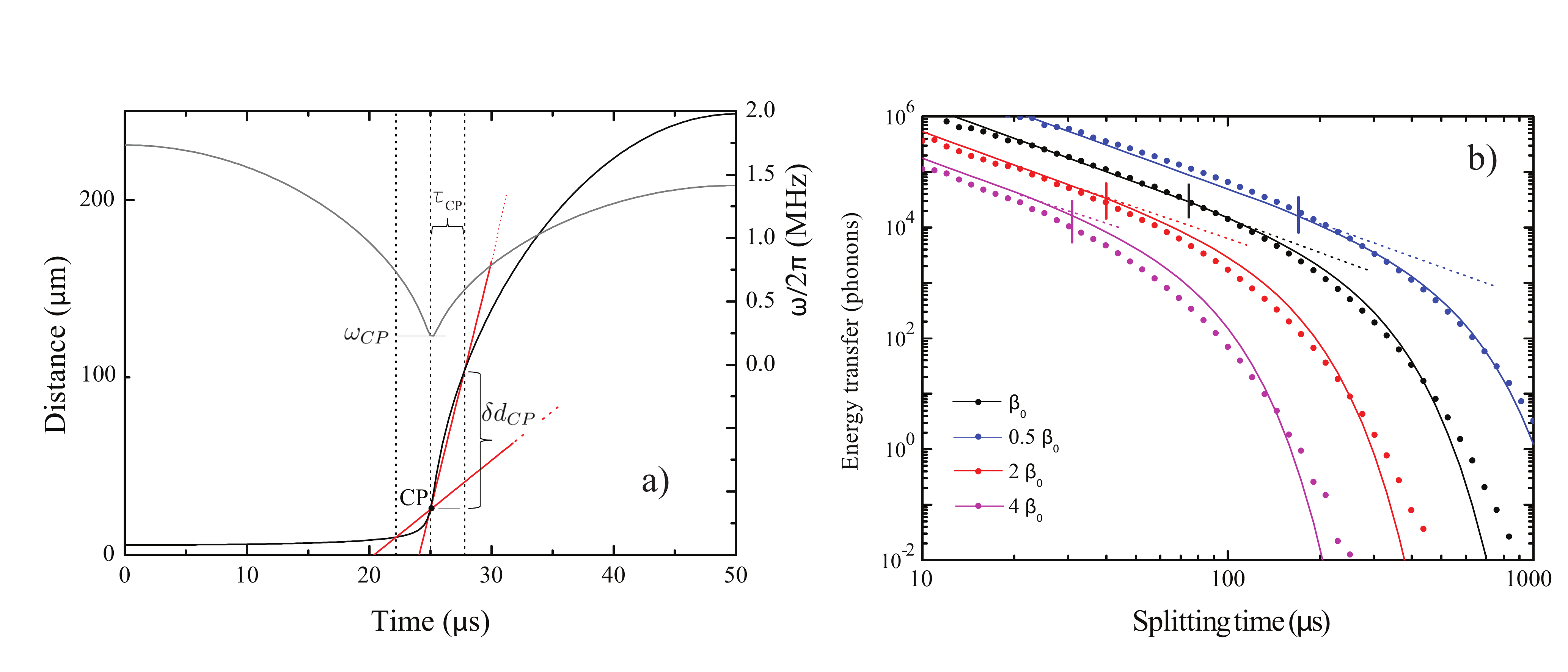}
\caption{Impulsive acceleration at the critical point. \textbf{a)} shows the equilibrium distance (black) versus time. The red lines depict the approximate slopes $\dot{d}_{CP}$ within time $\tau_{CP}$ before and beyond the CP. They illustrate how the impulsive displacement $\delta d_{CP}$ Eq. \ref{eq:deltadCP} is obtained from slope beyond the CP, and why the difference of the slopes, i.e. the second derivative $\ddot{d}_{CP}$, determines the onset of adiabaticity (see text). It is also shown how the trap frequency (gray) varies strongly during the CP trap period. \textbf{b)} compares the final excitation obtained from the simple approximation Eqs. \ref{eq:impulsiveapprox} (dashed), \ref{eq:impulsiveapproxadiab} (solid) to simulation results (dots). The onset of adiabaticity $\chi=1$, is marked with vertical bars. The calculations are carried out for a harmonic coefficients $\alpha(t)$ linearly varying around the CP, and different constant values for the quartic coefficient $\beta$.}
\label{fig:impulsiveapprox}
\end{figure}

A na\"{\i}ve approach towards crystal splitting is the linear interpolation between two voltage sets pertaining to a single well and a double well, leading to a constant variation rate of the harmonic coefficient $\alpha$. As this does not involve a dedicated control of the ion distance, it is equivalent to a rapid sweep across a structural transition of the ion crystal. This leads to an unfavorable power-law scaling of the energy transfer with respect to the sweep time \cite{ulm2013observation}, which prevents attaining adiabaticity.\\
In the following, we derive an approximation for the energy transfer, assuming the variation of $\alpha$ around the CP to be uniform. We consider the energy transfer to be caused by impulsive displacement: At the CP, the equilibrium distance changes most rapidly, while the confinement - and therefore the restoring forces - are reduced. Fig. \ref{fig:impulsiveapprox} a) shows that the situation corresponds to a harmonic oscillator which is suddenly dragged at uniform speed, causing displacement and therefore a gain in potential energy. Within the characteristic timescale set by half a the trap oscillation cycle $\tau_{CP}=\pi/\omega_{CP}$, this yields the displacement:
\bea
\delta d_{CP} &\approx & \dot{d}_{CP} \;\xi\; \tau_{CP}/2\\
						  &\approx & \left.\frac{\partial d}{\partial\alpha}\right\vert_{CP} \dot{\alpha}_{CP} \;\xi\; \tau_{CP}/2\\
						  &\approx & \left(\beta_{CP}^4~\kappa\right)^{-1/5} \dot{\alpha}_{CP} \;\xi\; \tau_{CP}/2,
\label{eq:deltadCP}
\eea
where Eq. \ref{eq:dpert} was used in the last line. The factor $\xi$ accounts for the fact that the trap frequency increases beyond the CP, such that the restoring forces set in before $\tau_{CP}$ and the resulting displacement is reduced. This sudden displacement mechanism is sketched in Fig. \ref{fig:impulsiveapprox} a). The potential energy of an ion is consequently increased by
\bea
\delta E &= & \frac{1}{2}m\omega_{CP}^2 \left( \delta d_{CP}/2\right)^2 \\
				 &= & \frac{\pi^2}{8} \;\xi^2 \;m \left(\beta_{CP}^4 ~\kappa\right)^{-2/5} \dot{\alpha}^2_{CP},
\label{eq:impulsiveapprox}
\eea
which serves as an approximation of the final energy transfer.\\
For a sufficiently small $\vert\dot{\alpha}\vert_{CP}$, adiabaticity sets in and the energy transfer scales exponentially with the splitting time. The reason for this is that
the Coulomb repulsion serves to push the ions outwards, providing smooth variation of the equilibrium distance as compared to discontinuous behavior of the minima of the external potential, see Fig. \ref{fig:eqnposnstogether} b). It therefore leads to rapid, but continuous variation of the equilibrium positions with $\alpha$.
The onset of the adiabatic regime is identified by comparing displacement $\delta d_{CP}$ to the change of the equilibrium distance within $\tau_{CP}$ \textit{below} the CP (see Fig. \ref{fig:impulsiveapprox} a)), which means that the ion acceleration around the CP is sufficiently slow to prevent sudden displacement. We therefore compare the acceleration $\ddot{d}_{CP}$ to the reference acceleration $d_{CP}\omega_{CP}^2$, yielding the adiabaticity parameter
\bea
\chi&=&\frac{\ddot{d}_{CP}}{d_{CP}\omega_{CP}^2} \\
&=&\frac{4}{25}\frac{m}{3e}2^{-1/5}\beta_{CP}^{-9/5}\kappa^{-6/5}\dot{\alpha}_{CP}^2
\eea
In the adiabatic regime, $\chi<1$, the energy transfer is given by:
\be
\delta E'\approx \delta E \exp\left[c^2 \left(1-\frac{1}{\chi}\right)\right]
\label{eq:impulsiveapproxadiab}
\ee
Numerical simulations are carried out for different constant values for $\beta$ and a linear variation of $\alpha$ around the CP. The results are shown in Fig. \ref{fig:impulsiveapprox} b). It can be seen that the approximations Eqs. \ref{eq:impulsiveapprox},\ref{eq:impulsiveapproxadiab} hold over a wide range of splitting times and quartic coefficients, and that large energy transfers in the regime of 10$^4$-10$^6$ phonons are readily obtained. The simulations yield a value of $\xi^2\approx$0.1.
We conclude that in this regime, the energy transfer depends only on the ion mass, the variation rate of $\alpha$  and the quartic confinement at the CP. As can be seen from the simulation results, still large energy transfers are obtained at the onset of adiabaticity, such that splitting at energy transfers on the single phonon level would require splitting times on the order of several hundreds of $\mu$s.\\
As we will show in further sections, this problem can be overcome using ramps that ensure a small ion acceleration $\ddot{d}_{CP}$ at the CP. Note that
\be
\ddot{d}_{CP}=\left.\frac{\partial^2 d}{\partial\alpha^2}\right\vert_{CP}\dot{\alpha}_{CP}^2+\left.\frac{\partial d}{\partial\alpha}\right\vert_{CP}\ddot{\alpha}_{CP}.
\label{eq:ddotdCP}
\ee
For sufficiently uniform variation of $\alpha$, the second term can generally be neglected, such that by using Eq. \ref{eq:dpert}, we obtain
\be
\ddot{d}_{CP}=\frac{2}{25}\left(\frac{4}{\beta_{CP}^7\kappa^3}\right)^{1/5}\dot{\alpha}_{CP}^2.
\label{eq:ddotdCPapprox}
\ee
Thus, the energy transfer can be reduced by ensuring a small variation rate of $\alpha$ at the CP.

\subsection{Uncompensated potential tilt}
\label{sec:uncomptilt}
A residual static force along the trap axis, expressed by the coefficient $\gamma'$ in Eq. \ref{eq:gammadef}, can originate from stray charges, laser induced charging of the trap \cite{HARLANDER2010}, trap geometry imperfections or residual ponderomotive forces along the trap axis.
The behavior of the equilibrium positions in the presence of an imperfectly compensated tilt, shown in Fig. \ref{fig:eqnposnstogether}, reveals a discontinuity for the critical $\tilde{\gamma}$, leading to diverging acceleration. The divergence of the acceleration impedes us to perform the splitting process adiabatically for $\vert\gamma\vert\lesssim\tilde{\gamma}$, i.e. the voltages can not be changed sufficiently slow to suppress motional excitation. Thus, one might encounter the situation that the tilt is sufficiently well compensated to allow for splitting, but sufficiently low excitations cannot be obtained irrespectively of the splitting time and other control parameters. For small tilt parameters, $\vert\gamma\vert\ll\tilde{\gamma}$, we can employ the perturbative expressions Eqs. \ref{eq:dpert}, \ref{eq:x0pert} of the equilibrium positions to obtain
\be
\frac{\partial^2 x_{R,L}}{\partial\alpha^2}=\frac{\partial^2 x_0}{\partial\alpha^2}\pm\frac{1}{2}\frac{\partial^2 d}{\partial\alpha^2}=\gamma \frac{52\cdot2^{4/5}}{675\beta^{9/5} \kappa^{6/5}}\pm\frac{2}{25}\left(\frac{4}{\beta^7\kappa^3}\right)^{1/5}
\ee
We can estimate the tilt parameter at which the acceleration of one of the ions is twofold compared to the tilt-free case determined by Eq. \ref{eq:ddotdCPapprox} to be about 67\% of the critical tilt $\tilde{\gamma}$. Due to the divergence of the acceleration at $\tilde{\gamma}$, we can expect the actual acceleration at this tilt value to be substantially larger, we thus conclude that a residual tilt $\vert\gamma\vert\ll\tilde{\gamma}$ is required to realize crystal splitting at low motional excitation. A possible experimental scheme for this has been demonstrated in \cite{EBLE2010}: The separation process is performed on a slow (second) timescale under continuous Doppler cooling and detection. The ion positions are extracted from the camera image, and a deviation of the center-of-mass from the initial value is restored by automatic adjustment of the outer electrode differential voltage $\Delta U_O$.

\subsection{Anomalous heating at the critical point}
\label{sec:anomheatingatCP}
Microstructured ion traps exhibit \textit{anomalous heating}, i.e. the mean phonon number increases due to thermalization with the electrodes at a timescale much faster than predicted by the assumption that only Johnson-Nyquist noise is present \cite{BROWNNUTT2013}. This process can be modeled as $\dot{\bar{n}}=\Gamma_h$, with the heating rate $\Gamma_h(\omega)= S_E(\omega) e^2 /4m\hbar\omega$
where the spectral electric-field-noise density $S_E$ depends on the trap frequency $\omega$. A polynomial decrease $S_E\propto\omega^{-a}$ is often assumed, where experimentally determined values for the exponent $a$ range from 0.5 to 2.5. Additionally, peaked features might arise in the noise spectrum which are caused by technical sources. Moreover, the absolute values of the heating rates strongly depend on the properties of the electrode surfaces. Typical values at trap frequencies in the 1~MHz regime range from 0.1 to tens of phonons per millisecond. As the trap frequency is strongly decreased around the CP, we can expect a significant amount of excess energy after the splitting caused by anomalous heating, increasing for longer splitting durations. We model this contribution by integrating over a time dependent heating rate:
\be
\Delta \bar{n}_{th}=\int_0^T \Gamma_h\left(\omega(t)\right) dt.
 \label{eq:avheating}
\ee
For the simulations that follow we will employ an experimentally determined relation for trap A (Sec. \ref{sec:geom}) which is $\Gamma_h(\omega)\approx 6.3\cdot\left(\omega/2\pi  \textrm{MHz}\right)^{-1.81} ms^{-1}$. This does not depend on the geometry of trap A but on the properties of our trap apparatus.\\
In the case of imperfect control of the ion distance around the CP, Sec. \ref{sec:accatCP}, or in the presence of an uncompensated tilt, Sec. \ref{sec:uncomptilt}, one will attempt to reduce the motional excitation by splitting very slowly. This might however be unsuccessful as anomalous heating will strongly contribute to the energy gain at large splitting times. Experimental procedures for ensuring a sufficient degree of control are therefore ultimately required.

\section{Voltage ramps}
\label{sec:voltageramps}
In this section we explain our scheme for designing voltage ramps for the splitting process. Our intention is to provide a scheme which can be applied any given trap geometry. We do explicitly not rely on the precise knowledge of the electrostatic trap potentials, but rather on quantities which can be measured with reasonable effort. Furthermore, we describe how a single voltage level can be used as a tuning parameter to achieve the optimum result. Our scheme assumes that the tilt potential is perfectly compensated, $\gamma=0$. We proceed as follows: We first describe how the segment voltages are supposed to vary with the harmonicity parameter $\alpha$, where we simply fix voltage levels on a small set of mesh points. We then show how this is used in conjunction with a chosen distance-versus-time and available distance-versus-$\alpha$ information to obtain time-domain voltage ramps which can be employed in the experiment.

\subsection{Static voltage sets}
\label{sec:staticramps}
The calculation of suitable voltage ramps relies on the signs and on the magnitude ordering of the geometry parameters. In Table \ref{tab:trapparams} we list values for several different microstructured traps. We assume that any reasonable segmented trap geometry will exhibit similar characteristics. From the results of Sec. \ref{sec:Intricacies}, it is clear that we desire a large positive value of $\beta_{CP}$. We assume that the voltages which can be applied to the segments are limited by hardware constraints to the symmetric maximum/minimum values $\pm U_{lim}$. To achieve the largest possible $\beta$ at the CP, we begin the splitting protocol by ramping the $O$ segments to $+U_{lim}$, keep them at constant bias during around the CP, and ramp them back to zero bias after the splitting.\\
The CP is defined by the condition $\alpha=0$, which is accomplished by suitable voltages $U_{C,S}$. This leaves one degree of freedom, which can be eliminated by maximizing $\beta_{CP}$. We solve Eq. \ref{eq:alphadef} for $U_C$ :
\be
U_C=\frac{1}{\alpha_C}\left(\alpha-\alpha_O U_O-\alpha_S U_S\right). \label{eq:UCfromUS}
\ee
The largest possible $\beta_{CP}$ is then given by inserting this result into Eq. \ref{eq:betadef} and setting $U_O^{(CP)}=+U_{lim},U_S^{(CP)}=-U_{lim}$:
\be
\max_{U_C,U_S}\beta_{CP}=\left(\beta_O+\frac{\beta_C}{\alpha_C}\alpha_S-\beta_S-\frac{\beta_C}{\alpha_C}\alpha_O\right)U_{lim}
\label{eq:betaMax}
\ee
Static splitting voltage sets are obtained by fixing the initial, CP and final voltage configurations and interpolating between these. The procedure consists of the following steps:
\begin{enumerate}
\item
Determine the initial $\alpha_i>0$ from Eq. \ref{eq:alphadef} using the initial voltages $U_C^{(i)}<0$~V, $U_S^{(i)}=U_O^{(i)}=0$~V.
\item
Choose the voltages at the CP such that the maximum $\beta_{CP}$ is attained, by setting $U_O^{(CP)}=+U_{lim},U_S^{(CP)}=-U_{lim}$ and $U_C^{(CP)}$  from Eq. \ref{eq:UCfromUS} for $\alpha=0$. \footnote{If the geometry parameters are such that $U_C^{(CP)}$ exceeds $\pm U_{lim}$, set $U_C^{(CP)}=-U_{lim}$ and obtain $U_S^{(CP)}$ solving Eq. \ref{eq:alphadef} for $U_S$ rather than $U_C$.} \footnote{If the magnitude of $U_S^{(CP)}$ is chosen smaller than $U_{lim}$, this  leads to smaller values of $\beta_{CP}$ and a larger ion separation at the CP. This offers the possibility for well-controlled studies of the dependence of the splitting process on the quartic confinement at the CP.}.
\item
Determine the desired final voltages.  We choose $U_C^{(f)}=0$~V, $U_S^{(f)}=U_S^{(CP)}=-U_{lim}$ and $U_O^{(f)}=0$~V. This choice is convenient when $U_C^{(i)}\approx -U_{lim}$ and ensures that the ions are finally kept close to the respective centers of the $S$ segments with a trap frequency similar to the initial one. Obtain $\alpha_f$ from Eq. \ref{eq:alphadef}.
\item
For approaching the CP, $\alpha_i\geq\alpha >0$, set
\be
U_S(\alpha) = \left(1-\frac{\alpha}{\alpha_i}\right) U_S^{(CP)}
\label{eq:USbelowCP}
\ee
and
\be
U_O(\alpha) = \cases{2\left(1-\frac{\alpha}{\alpha_i}\right) U_{lim}\qquad\alpha> \frac{\alpha_i}{2} \\ U_{lim}\qquad \ \ \alpha \leq \frac{\alpha_i}{2}\\}
\label{eq:UObelowCP}
\ee
and obtain $U_C(\alpha)$ from Eq. \ref{eq:UCfromUS}.
\item
Beyond the CP, $0\geq\alpha\geq \alpha_f$, set
\be
U_S(\alpha)=-U_{lim}
\label{eq:USaboveCP}
\ee
and
\be
U_O(\alpha) = \cases{U_{lim}\qquad \alpha > \frac{\alpha_f}{2} \\
    2\left(1-\frac{\alpha}{\alpha_f}\right) U_{lim} \ \ \ \ \alpha \leq \frac{\alpha_f}{2} \\}
\label{eq:UObeyondCP}
\ee
and obtain $U_C(\alpha)$ from Eq. \ref{eq:UCfromUS}.
\end{enumerate}

\begin{figure}[ht!]
\centering
\includegraphics[width=0.75\textwidth]{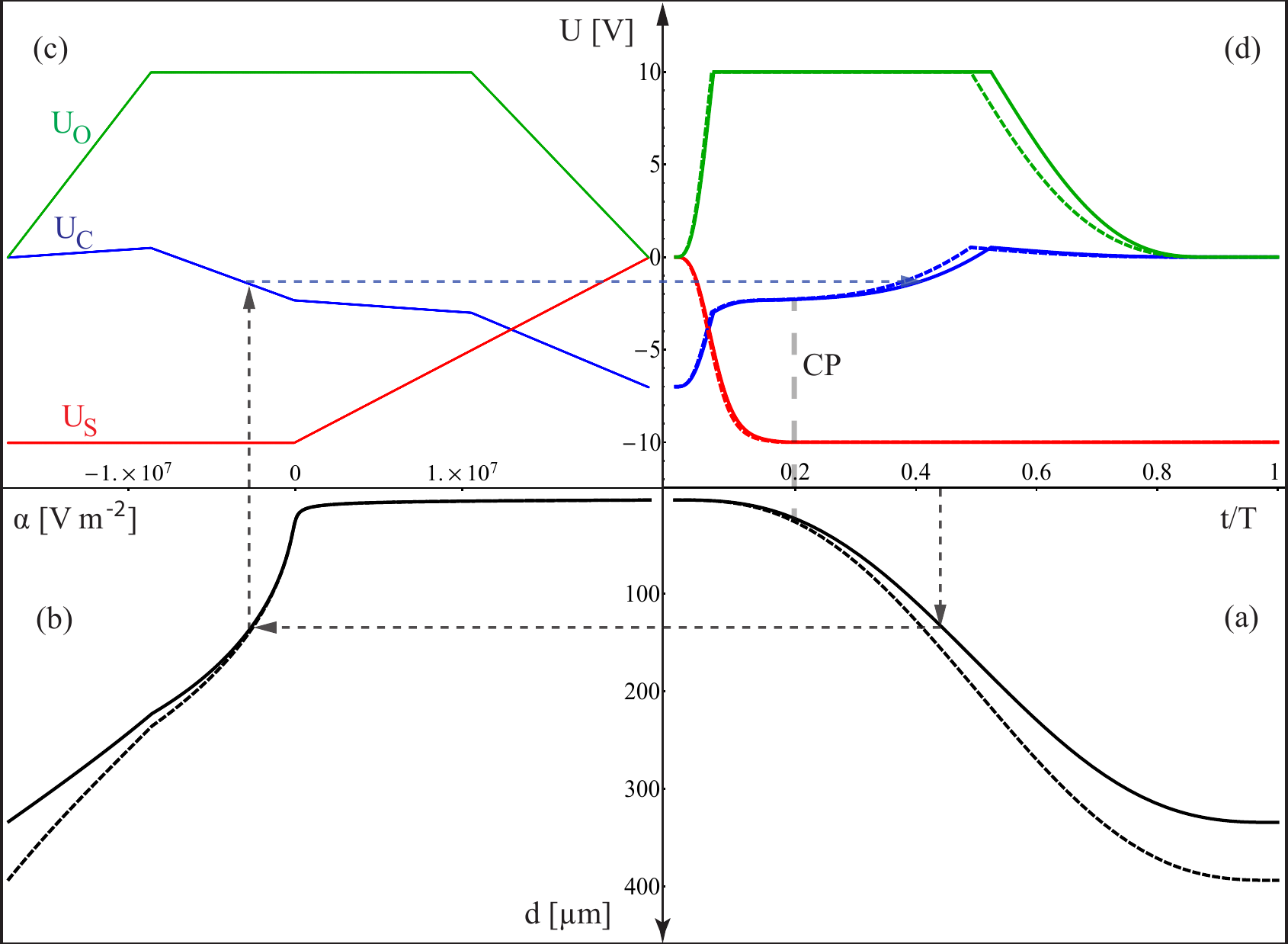}
\caption{Voltage ramp transfer to the time domain: A predefined time-to-distance function $d(t)$ shown in panel \textbf{a)} is used in conjunction with $\alpha$-to-distance information $\alpha(d)$ shown in \textbf{b)} to determine the time-dependent electrode voltages $U_n(t)$ using the static voltage sets $U_n(\alpha)$ from panel \textbf{c)}. The resulting ramps $U_n(t)$ are shown in \textbf{d)}. The dashed curves are corresponding to the case when the voltage ramps are calculated according to the presented method, but realistic trap potentials from simulations are used to determine $d_f$ and $d(\alpha)$. The dashed arrows exemplify how a specific value $U_C$ is obtained.}
\label{fig:timedomainvoltageramps}
\end{figure}

\subsection{Time domain ramps}
\label{sec:timeramps}

We now show how to design suitable time-domain voltage ramps $U_n(t)$ that will assure well-controlled splitting. It has been shown in Sec. \ref{sec:accatCP} that a small value of the acceleration at the CP, $\ddot{d}_{CP}$, is required for achieving a low energy transfer. This in turn is guaranteed by well-controlled variation of of the distance $d(t)$ throughout the splitting process.  As $d(\alpha)$ is monotonically decreasing with $\alpha$, it can be inverted to obtain $\alpha(d)$ which is used to compute the final voltage ramp as $U_n(\alpha(d(t)))$ (see Fig. \ref{fig:timedomainvoltageramps}.).\\
Possible choices for $d(t)$ are a sine-squared ramp
\be
d(t)=d_i+\left(d_f-d_i\right)\sin^2\left(\frac{\pi t}{2T}\right)
\ee
or a polynomial ramp
\be
d(t)=d_i+\left(d_f-d_i\right)\left(-10\frac{t^3}{T^3}+15\frac{t^4}{T^4}-6\frac{t^5}{T^5}\right)
\ee
Both ramps fulfill $d(0)=d_i,d(T)=d_f,\dot{d}(0)=\dot{d}(T)=0$. The polynomial ramp, used in the following, additionally fulfills $\ddot{d}(0)=\ddot{d}(T)=0$, while the second derivative of the sine-squared ramp displays discontinuities. However, these features presumably play no role in experiments, as the voltage ramps are generally subject to discretization and filtering. Different methods can be employed for the determination of $d(\alpha)$:
\begin{itemize}
\item
The equilibrium distance can be computed by employing realistic trap potentials from simulation data, using the voltage configuration pertaining to a given $\alpha$ as determined by the static voltage sets $U_n(\alpha)$. This method requires the simulated potentials to match the actual trap potential with great precision.
\item
The equilibrium distance can be computed using values from calibration measurements for the coefficients $\alpha_n,\beta_n$. This circumvents the need for simulations and accounts for parameter drifts. It yields only valid values for distances which are small compared to the electrode width, however we will show in Sec. \ref{sec:simu} that this procedure yields useful voltage ramps.
\item
Ion distances can be measured by imaging the ion crystal on a camera, while voltages configurations for decreasing $\alpha$ values are applied. This is the most direct method, and it benefits from the availability of a precise gauge of imaging magnification from measurements of the trap frequency.
\end{itemize}

\section{Simulation results}
\label{sec:simu}
In order to analyze the sensitivity of the splitting process and the performance of our ramp design protocol, we numerically solve the classical equations of motion. For the time- and energy-scales and potential shapes under consideration, we expect quantum effects to play no significant role. For the case of single-ion shuttling, the occurrence of quantum effects is thoroughly discussed in Ref. \cite{furst2013controlling}.\\
We perform the simulations using either the Taylor approximation of the potentials or the realistic potentials from electrostatic simulations \cite{SINGER2010} for trap A, which is similar to that described in Ref. \cite{SCHULZ2008}. The voltage ramps $U_i(t)$ are used in conjunction with the potentials to yield the equations of motion for the ion positions $x_1<x_2$. Employing the Taylor approximation potential Eq. \ref{eq:taylorpotential}, these read
\be
-m\ddot{x}_{1,2}=4\beta(t) x_{1,2}^3+2\alpha(t) x_{1,2}+\gamma\pm\frac{\kappa}{(x_2-x_1)^2},
\label{eq:eqofmotion1}
\ee
where the coefficients are given by using the voltage ramps in Eqs. \ref{eq:alphadef}, \ref{eq:betadef},\ref{eq:gammadef}. For realistic trap potentials, we obtain
\be
-m\ddot{x}_{1,2}=\sum_{n=C,S,O} U_n(t) \left. \frac{d\phi_n}{dx}\right|_{x_{1,2}}\pm\frac{\kappa}{(x_2-x_1)^2}\\
\label{eq:eqofmotion2}
\ee
The possibility to perform the simulations with approximate and realistic potentials serves the purpose of verifying the performance of the voltage ramps. These are determined purely by trap properties around the CP, which are conveniently accessible by measurements. More precisely, the time-domain voltage ramps are based on a $d(\alpha)$ dependency given by the Taylor approximation potential according to Fig. \ref{fig:timedomainvoltageramps}, while the resulting energy transfer pertaining to these ramps can be obtained from simulations using realistic potentials.\\
Note that a nonzero tilt can be present in the simulations based on the realistic potentials by summing separately over electrodes $O_L$ and $O_R$ and adding the differential voltage $\pm\Delta U_O$ given by $\gamma/\gamma_O$ accordingly. The calculations presented here employ the mass of $^{40}$Ca$^+$ ions which we use in our experiments, and all simulations were performed for a limiting voltage range  $U_{lim}=10$~V.\\
Eqs. \ref{eq:eqofmotion1} or \ref{eq:eqofmotion2} are solved numerically using the  \textit{NDSolve} package from \textit{Mathematica}, with the ions starting at rest. The final oscillation of each ion around its equilibrium position is analyzed and yields the energy transfer expressed as the mean phonon number $\bar{n}=\Delta E/\hbar\omega_f$. We distinguish several regimes of laser-ion interaction: i) If the vibrational excitation becomes so large that the average Doppler shift per oscillation cycle exceeds the natural linewidth of a cycling transition, ion detection by counting resonance fluorescence photons will be impaired. ii) Measurement of the energy transfer i.e. by probing on a stimulated Raman transition \cite{WALTHER2012} typically requires mean phonon numbers below about 300. iii) The Lamb-Dicke regime of laser-ion interaction, where coherent dynamics on resolved sidebands can be driven \cite{leibfried2003experimental} is typically attained below about 10 phonons. The borders between these regimes depend on the trap frequency, ion mass and the specific atomic transitions to be driven, thus the regimes are indicated as broad gray bands in Fig. \ref{fig:nbarVsTf}. Note that if final excitations in the measurable regime are obtained, an electrical counter kick can be applied for bringing the oscillation to rest \cite{WALTHER2012}.\\

\subsection{Dependence on splitting time}
\begin{figure}[h!]
\centering
\includegraphics[width=13cm]{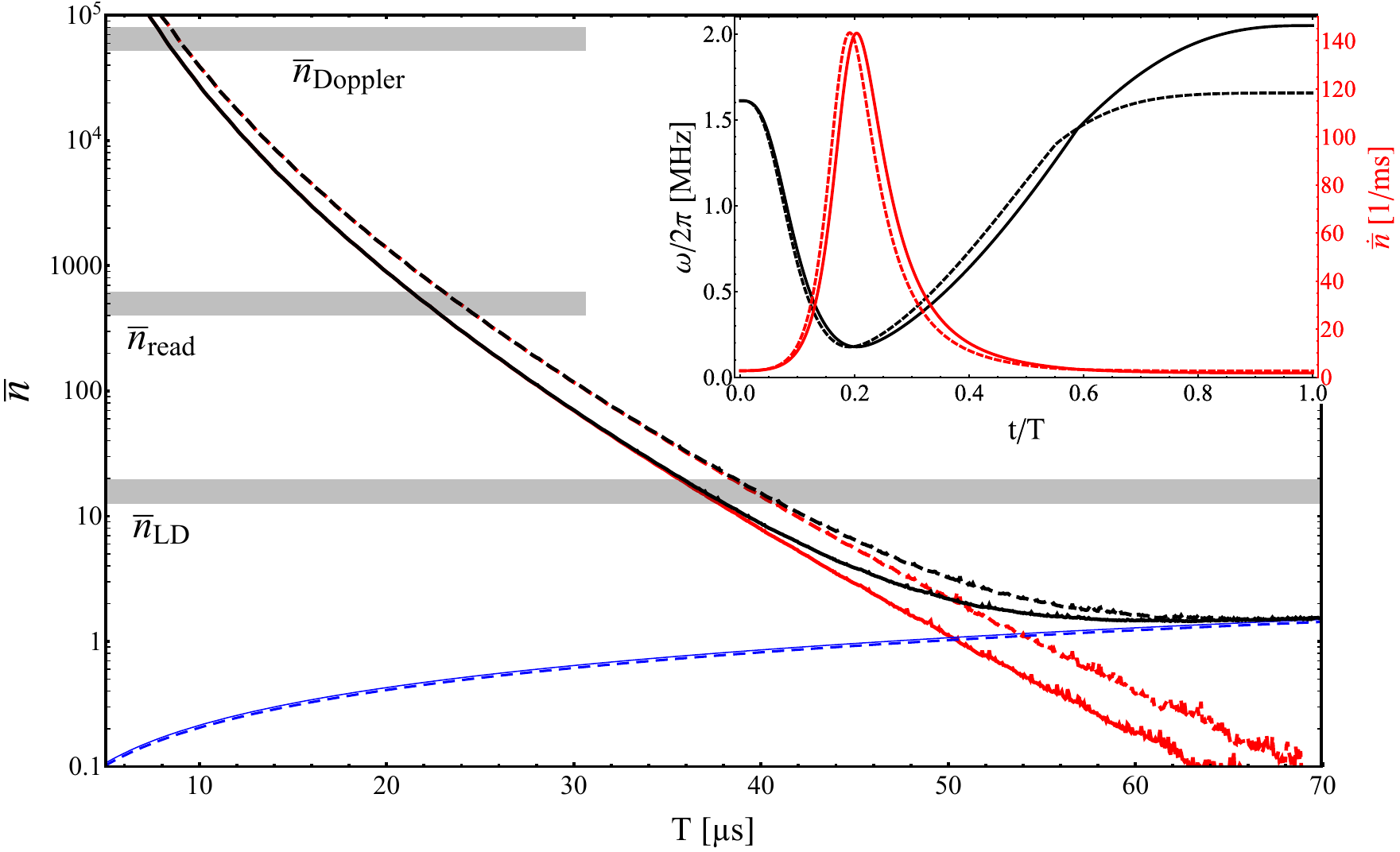}
\caption{Energy transfer versus splitting time: Oscillatory (red) and thermal excitation (blue), and the sum of both (black) versus the splitting duration $T$.  The solid lines correspond to the calculation using the Taylor approximation, the dashed lines correspond to the full potential calculation, see text. Grey bands seperate different regimes of laser-ion interaction, see text. The thermal excitation was deduced from experimental heating rate data according to Sec. \ref{sec:anomheatingatCP}. The inset shows the trap frequency (black) and the corresponding heating rate (red) as a function of normalized time during the splitting process.}
\label{fig:nbarVsTf}
\end{figure}

We first analyze the dependence of the energy transfer on the duration of the splitting process $T$, the result is shown in Fig. \ref{fig:nbarVsTf}. The calculation is carried out for the ideal case of perfectly compensated potential tilt. We see that the final excitation becomes sufficiently low to remain in the Lamb-Dicke regime for typical laser-ion interaction settings at times larger than about 40~$\mu s$, which clearly outperforms the na\"{i}ve approach of voltage interpolation from Sec. \ref{sec:accatCP}.\\
We also take into account increased anomalous heating around the CP by employing the averaged heating rate according to Eq. \ref{eq:avheating}. We see that for our specific heating rates, the limit of about one phonon per ion can not be overcome, but as the anomalous heating contribution is scaling as $1/T$, the splitting result becomes rather  insensitive with respect to the precise choice of the $T$ beyond  $T=$~50~$\mu s$. \\
The simulation results verify our approach of calculating the voltage ramps using the Taylor approximated potentials. One recognizes that the resulting energy transfer in this case is larger by a factor of about two throughout the entire range of splitting durations. As can be seen from Fig. \ref{fig:timedomainvoltageramps}, this is due to the fact that the Taylor expansion leads to an incorrect voltage set pertaining to the CP, which in turn leads to uncontrolled acceleration as explained in Sec. \ref{sec:accatCP}. The discrepancy becomes irrelevant for splitting times larger than  $T=$~60~$\mu$s. At around $60$ to $70~\mu$s the oscillatory excitation becomes smaller than $\bar{n}=0.1$, corresponding to the limit we can currently resolve in our experiment. The slight inaccuracy for low phonon numbers is due to numerical artifacts. Even lower energy transfers at shorter $T$ could possibly be achieved by ramp engineering, i.e. by the application of shortcut-to-adiabaticity approaches \cite{furst2013controlling,palmero2013fast}.

\subsection{Sensitivity analysis}

\begin{figure}[h!]
\centering
\includegraphics[width=15cm]{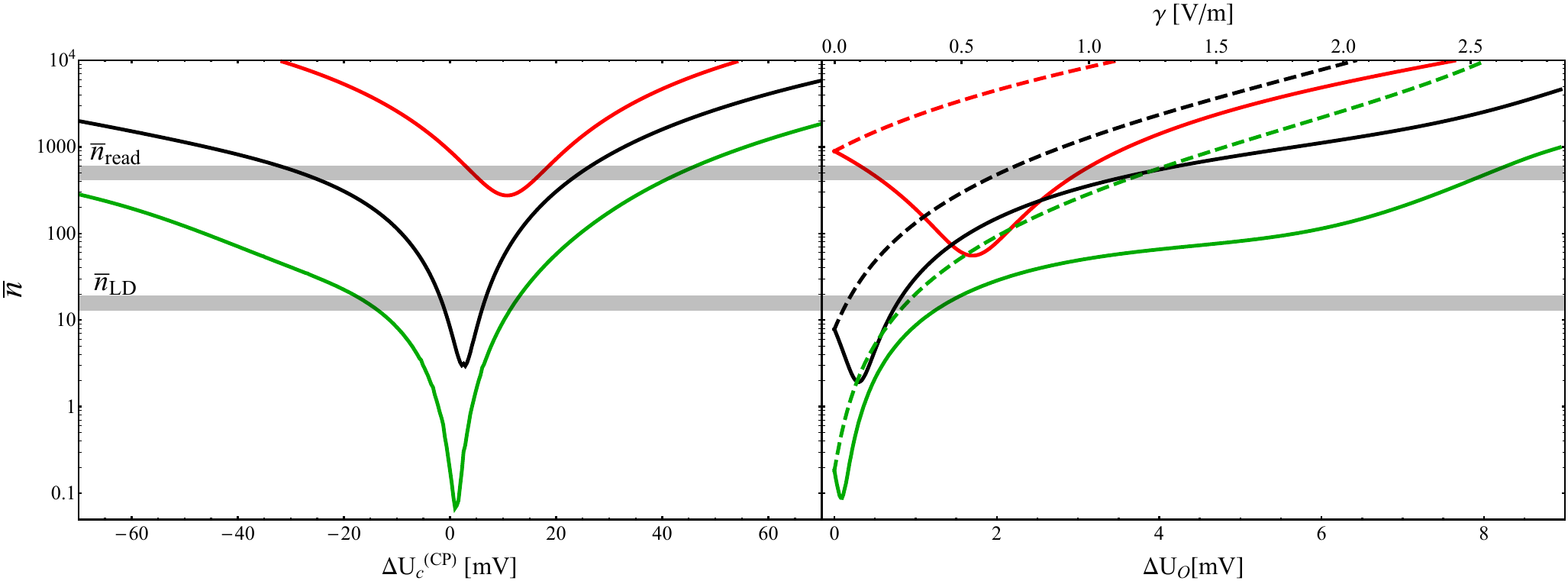}
\caption{Mean coherent excitation as a function of the offset voltage at the center segment at the CP (a) and the tilt force $\gamma$  (b). The tilt voltage $+\Delta U_O$ is applied to the right outer segment and $-\Delta U_O$ is applied to the left outer segment. The mean phonon number for the right ion is depicted by dashed lines and by solid lines for the left ion. The curves correspond to different splitting times: $T = 60\mu s$ (green), $T = 40\mu s$ (black),  $T = 20\mu s$ (red). The critical tilt is at $\tilde{\gamma}=3~$V/m.}
\label{fig:nbarVsGamma}
\end{figure}

Two crucial parameters for the splitting operation are the offset voltage at the CP $\Delta U_C^{(CP)}$ and the potential tilt $\gamma$. Small variations of these parameters lead to strong coherent excitations as shown in Fig. \ref{fig:nbarVsGamma}.\\
The CP voltage offset $\Delta U_C^{(CP)}$ serves both for modeling and compensation of inaccuracies of the trap potentials, leading to a wrongly determined CP voltage configuration and therefore to increased acceleration. It is implemented into the simulations by just adding it to $U_C^{(CP)}$ as determined by Eq. \ref{eq:UCfromUS} in the calculation of the static voltage sets. We see that even for sufficiently slow splitting, the Lamb-Dicke regime can only be attained if this voltage offset, and therefore the CP voltages in general, are correct within a window of about 20~mV, on the other hand it becomes clear that this voltage serves as convenient fine tuning parameter. The minimum excitation does not occur at $\Delta U_C^{(CP)}=0$, but is slightly shifted to positive values.\\
This can be understood by considering that $\vert\dot{\alpha}\vert_{CP}$ is increased for any $\Delta U_C^{(CP)} \neq 0$, but  $\ddot{\alpha}_{CP}$ is decreased for $\Delta U_C^{(CP)} >0$. With $\partial d/\partial\alpha$, the second term in Eq. \ref{eq:ddotdCP} leads to a reduced total acceleration for small positive  $\Delta U_C^{(CP)}$. Larger values again lead to increased acceleration because of a smaller $\beta_{CP}$ value. All other calculations in this work are done using $\Delta U_C^{(CP)}=0$.\\
For the case of an uncompensated tilt $\gamma'$, we observe an even stronger dependence of the energy transfer. Fine tuning of the voltage difference on the outer segments $\Delta U_O$ on the sub-mV level is required to reach the single phonon regime. Moreover, we observe that moderate uncompensated potential tilts reduce the energy transfer to one of the ions, as its CP acceleration is reduced by a more smooth $x(\alpha)$ dependence. This might be of interest for specific applications where only the energy transfer to one of the ions is of importance.

\subsection{Dependence on the limiting voltage}

\begin{figure}[h!]
\centering
\includegraphics[width=10cm]{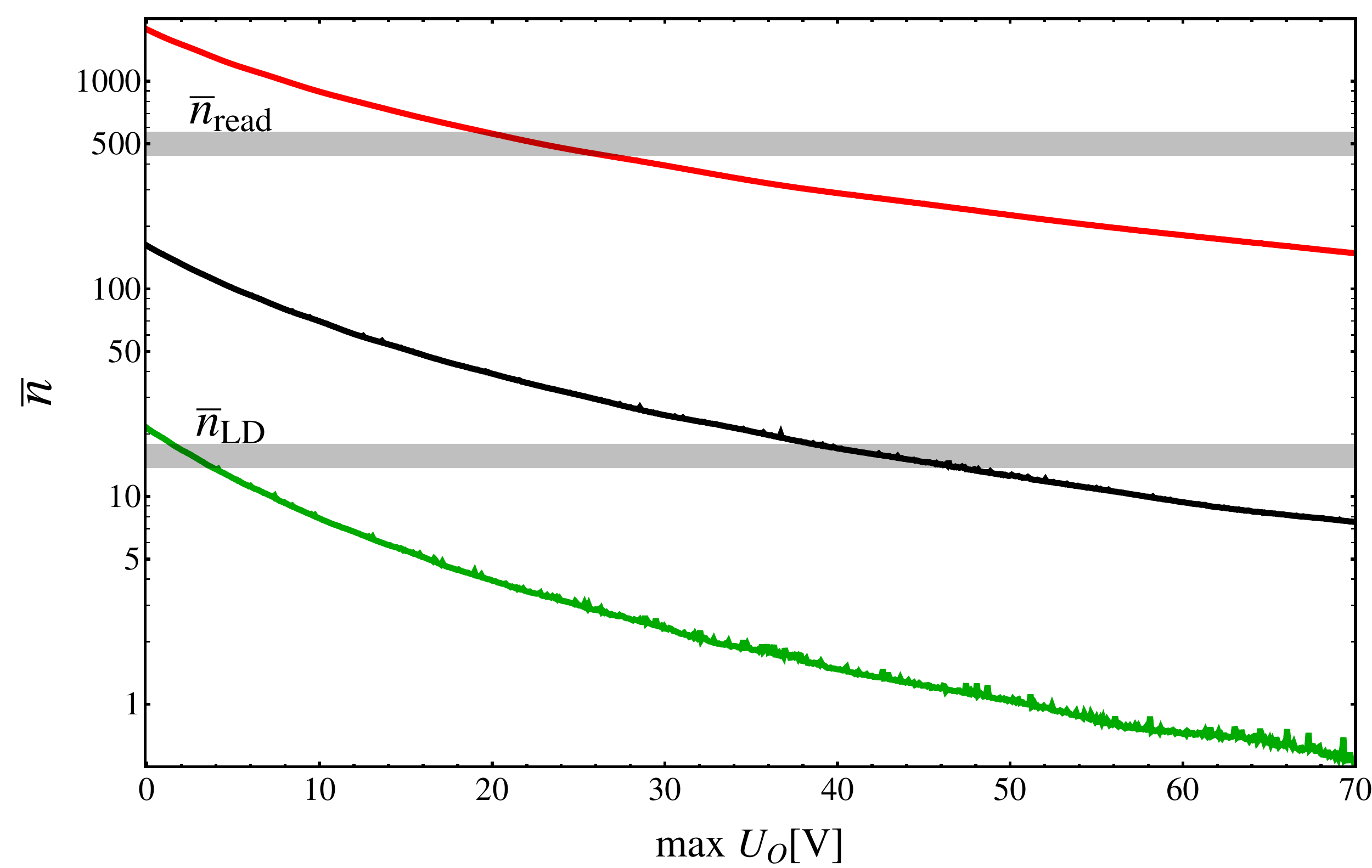}
\caption{Dependence on the voltage limit: Oscillatory excitation as a function of the maximum voltage on the outer segments with all other limiting voltages remaining unchanged. The curves correspond to different splitting times: $T = 40\mu s$ (green), $T = 30\mu s$ (black),  $T = 20\mu s$ (red). }
\label{fig:outerTune}
\end{figure}

Finally we study the dependence of the energy transfer on the limiting voltage $U_{lim}$. We find that by increasing the voltage limit, beyond $U_{lim}=10$~V used so far, we can obtain lower coherent excitations as shown in Fig. \ref{fig:outerTune}. For this simulation, only the maximum voltage on the outer segments (max $U_O$) is increased and all other limits remain unchanged. We infer that by increasing the voltage limit on these electrodes up to about $50$~V, one can reduce the mean phonon number by a factor of $\approx 8$ for $T=60\mu$s. For lower splitting durations the enhancing factor becomes slightly smaller.

\section{Trap geometry optimization}
\label{sec:geom}
\begin{figure}[h!]
\centering
\includegraphics[width=\textwidth]{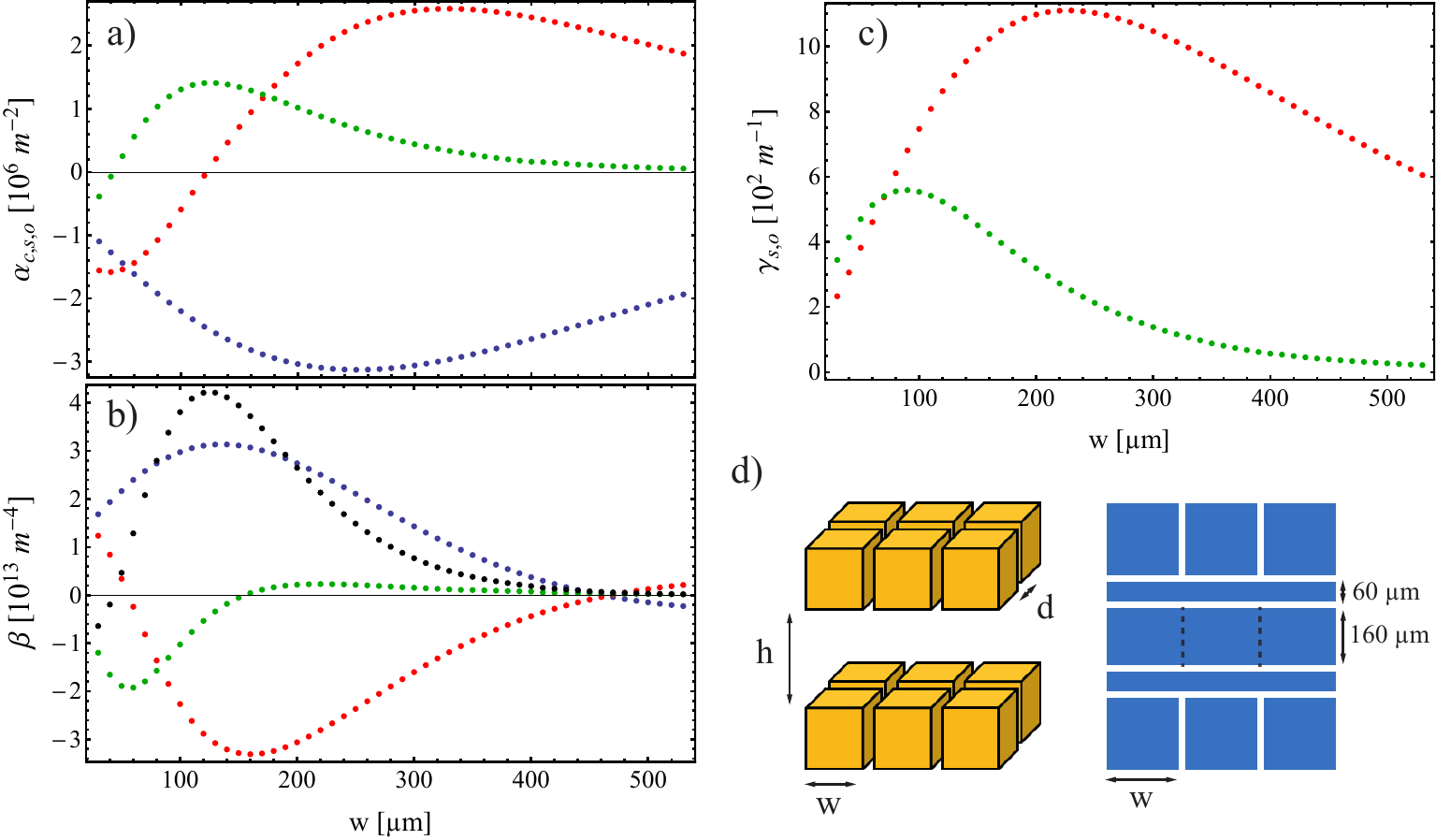}
\caption{Calculated geometry parameters $\alpha_n,\beta_n,\gamma_n$ and the maximum $\beta_{CP}$ at the critical point for a linear segmented Paul trap with dimensions $h=400~\mu$m, $d=250~\mu$m as a function of the segment width $w$. The color code is as above: blue - C, red - S, green - O. The limiting voltage for the electrodes is $U_{lim}=10V$. }
\label{fig:4erPlotGeomPar}
\end{figure}

We have been showing in Sec. \ref{sec:Intricacies} that the outcome of a crystal splitting operation is strongly determined by magnitude of the quartic confinement coefficient at the CP $\beta_{CP}$ from Eq. \ref{eq:betaMax}. We thus investigate the effect of the trap geometry on the coefficients $\alpha_n,\beta_n,\gamma_n$ from Eqs. \ref{eq:alphabetafromPhi}. We calculate the realistic potentials from electrostatic simulations \cite{SINGER2010} to infer the geometry parameters according to Eq. \ref{eq:alphabetafromPhi}. In particular, six different traps designs were studied, four of which are three-dimensional and two are surface-electrode traps. The results are shown in Tab. \ref{tab:trapparams}. The calculations are carried out for a generic simplified geometry shown in Fig. \ref{fig:4erPlotGeomPar} d), which is essentially determined by the segment width $w$, the slit height $h$ and the spacer thickness $d$ for the three-dimensional traps. Trap A ,B\cite{SCHULZ2008} and C\cite{BLAKESTAD2011} are similar segmented micro-structured ion traps . Trap B is subdivided into a loading region of larger geometry, B (wide), and a narrow processing region, B (narrow). The data for trap C pertains to a wedge segment of $w=100\mu$m surrounded by larger segments. Trap D is a segmented planar ion trap \cite{NARAYANAN2011}, the calculations are performed at a distance of $100~\mu$m between the ion and the surface. Trap D2 is a planar ion trap featuring a segmented ground plane, otherwise identical to trap D. Trap A was used for all simulations in section \ref{sec:simu}.\\

\begin{table}[htb]
\centering
    \begin{tabular}{| l | l | l | l | l | l | l |l |}
    \hline
     Parameter & Unit & A &B (wide) & B (narrow) & C & D & D2 \\  \hline
     $w$ & $\mu$m & 200 & 250 & 125 & 100 &200 & 200\\ \hline
     $h$ & $\mu$m & 400 & 500 & 250 & 200 &- &-\\ \hline
     $d$ & $\mu$m & 250 & 125 & 125 & 250 &- &-\\ \hline
	 $\alpha_C$ & 10$^{6}$ m$^{-2}$ & -3.0 & -2.5  & -9.1  & -6.4 & -1.4 & -12.0\\  \hline
     $\beta_C$ & 10$^{13}$ m$^{-4}$ & 2.7 &  1.7 & 19.9 & 14.4 & 1.5 & -6.5\\  \hline
     $\alpha_S$ & 10$^{6}$ m$^{-2}$ & 1.7 & 1.7 & 6.2 &  4.7 & 0.9 & 10.7\\  \hline
     $\beta_S$ & 10$^{13}$ m$^{-4}$ & -3.0 & -1.9 & -22.1 & -14.7 & -1.7 &5.6\\  \hline
     $\gamma_S$ & 10$^{2}$ m$^{-1}$ & 11.0 & 9.3 & 19.2 & 21.6 & 4.1 &17.8\\  \hline
     $\alpha_O$ & 10$^{6}$ m$^{-2}$ & 1.0 & 0.6 & 2.3 & 1.6 & 0.4 &0.9\\  \hline
     $\beta_O$ & 10$^{13}$ m$^{-4}$ & 0.2 & 0.2 & 2.0 & 1.2 & 0.1 &0.8\\  \hline
     $\gamma_O$ & 10$^{2}$ m$^{-1}$ & 3.2 & 2.2 & 4.3 & 3.2 & 1.2 &2.2\\  \hline
     $\omega_{CP}/2\pi$ & MHz & 0.18 & 0.14 & 0.29 & 0.26 &0.14 &0.11\\  \hline
    \hline
    \end{tabular}
\caption{Comparison of trap geometry parameters for different linear segmented Paul traps. Letters A to D denote different traps which are operated at various institutes, see text. Note that $\gamma_C=0$ by definition. The trap frequency at the critical point is specified for $U_{lim}$=10V and $^{40}$Ca$^+$ ions.}
\label{tab:trapparams}
\end{table}

For trap A and B (wide) we calculate similar parameters, however the minimum trap frequency during the splitting is larger for trap A. Trap B (narrow) exhibits the highest minimum trap frequency of the six geometries as the total dimensions of this section of the trap are rather small. The wedge segment in trap C helps to increase the minimum trap frequency but choosing an overall smaller size seems to be a more favorable solution. The planar trap D has a similar minimum trap frequency as trap B (wide) and is also suitable for splitting ion crystals. The segmentation of the ground plane of this trap (D2) offers an enhanced $\alpha_C$, i.e. a large trap frequency. The calculations show however that for a segmentation of the center electrode, the potentials become more anharmonic and the Taylor approximation Eq. \ref{eq:taylorpotential} breaks down. Thus, the sign and magnitude ordering of the coefficients might be different from the other geometries, therefore the geometry parameters and the ion height above the surface should be carefully chosen to allow for successful splitting operations.\\

For trap A we calculated the geometry parameters for varying segment width $w$, the result is shown in Fig. \ref{fig:4erPlotGeomPar}.
We analyze the dependance of all potential coefficients on $w$ with parameters $h$ and $d$ held constant. For splitting operations the optimum segment width would be at about $w=125\mu$m, while the actual segment width of the trap is $w=200\mu$m. We could therefore obtain a roughly twofold increase of $\beta_{CP}$ bought at the expense of a reduced trap frequency for ion storage due to the reduced $\alpha_C$ coefficient. \\ \newline
Finally, we investigate the dependence of $\beta_{CP}$ on the overall trap geometry size. We therefore pick trap parameters $h$ and $d$ from the range of typical values and determine the optimum segment width $w$ for these. Defining the effective trap size $d_{eff}=\left(w^2+h^2+d^2\right)^{1/2}$, we find a scaling behavior of $\beta_{CP}\approx 2.2 \cdot 10^{24} V \cdot d_{eff}^{-4}$, i.e. the best attainable value for the quartic confinement coefficient scales as the inverse fourth power with the effective trap size, which is the similar to the presumed distance scaling law for anomalous heating \cite{BROWNNUTT2013}. We conclude that for a trap architecture aiming at shuttling-based scalable quantum information, the considerations presented here should be incorporated into the design process to facilitate crystal splitting operations.

\section{Conclusion}
\label{sec:conclusion}
We have pointed out the pitfalls for ion crystal splitting: Uncontrolled separation and uncompensated background fields lead to enhanced acceleration of the ions when the single well potential is transformed into a double well, which would require splitting times in the millisecond range to keep the motional excitation near the single phonon level. This in turn leads to strong anomalous heating due to the reduced confinement during the splitting process. We presented a framework to design voltage ramps which allow for coping with these problems. The scheme does only rely on measured calibration data which is obtained for the initial situation, where the ions are tightly confined in a single potential well. We carried out simulations, which elucidate the energy transfer mechanisms, and verify the performance of our scheme for the voltage ramp calculation. We showed that excitations near the single phonon level can be obtained for the specific trap apparatus we use. Furthermore, we analyzed the suitability of different trap geometries for ion crystal splitting by means of electrostatic simulations. We concluded that crystal splitting becomes easier for smaller trap structures, and that dedicated optimization of the geometry can be helpful. In future work, we envisage to analyze how crystal splitting can be performed on faster timescales by using shortcut-to-adiabaticity approaches, with an emphasis on robustness against experimental imperfections.\\

\section*{Acknowledgments}
We thank Ren\'{e} Gerritsma and Georg Jacob for proofreading the manuscript. This research was funded by the Office of the Director of National Intelligence (ODNI), Intelligence Advanced Research Projects Activity (IARPA), through the Army Research Office grant W911NF-10-1-0284. All statements of fact, opinion or conclusions contained herein are those of the authors and should not be construed as representing the official views or policies of IARPA, the ODNI, or the US Government. CTS acknowledges support from the German Federal Ministry for Education and Research (BMBF) via the Alexander von Humboldt Foundation.

\vspace{1cm}

\bibliographystyle{unsrt}
\bibliography{splitting}

\end{document}